# "Applying mathematical models in cloud computing: A survey"

Alexander NGENZI

*Department of Computer Science Engineering-School of Engineering*

*Jain University*

*yngenzi37@gmail.com*

**ABSTRACT**

*As more and more information on individuals and companies are placed in the cloud, concerns are beginning to grow about just how safe an environment it is. It is better to prevent security threats before they enter into the systems and there is no way how this can be prevented without knowing where they come from. The issue of resource allocation and revenue maximization is also equally important especially when it comes to cloud security. This brings about the necessity of different modelling techniques including but not limited; security threat, resource allocation and revenue maximization models. This survey paper will try to analyse security threats and risk mitigation in cloud computing. It gives introduction of how viral attack can invade the virtual machines on the cloud, discusses the top security threats and countermeasures by providing the viral threat modelling in virtual machines and risk mitigation. Resource allocation models and revenue maximization techniques are also discussed.*

## 1. Introduction

The majority of organizations in the commercial and government sectors now use digital Information Technology (IT) to store and process data that is sensitive in some way. Sensitive data ranges from individuals' confidential details to valuable intellectual property to market sensitive information or even state secrets. At the same time, the commercialization of the Internet in the mid-1990s has resulted in the Internet becoming the de facto electronic channel over which organizations now interact with each other. Even where systems are not directly connected to the Internet, there are often indirect channels being inadvertently created to reach apparently disconnected systems. The increase in connectivity has bought about new threats and that threat continues to evolve as connectivity evolves with developments such as mobile devices.

This special issue is intended to bring forth the recent advancements in the detection, modelling, monitoring, analysis and defense of various threats posed to sensitive data and security systems from unauthorized or other inappropriate access. Also, resource allocation has been our concern especially when need to post services and deliver them from the cloud. Nevertheless, revenue maximization is equally important for customer satisfaction. In this paper, we propose applying mathematical models in cloud computing to examine the impact of all mentioned above to distinguished customers . The phrase "Cloud" originates from the cloud symbol used by flow charts and diagrams to symbolize the Internet. The term Cloud Computing refers to both the applications delivered as services over the Internet and the servers and system software in the datacenters that provide those services. The virtual machines(VMs) on the cloud will be affected due to the sharing of resources among themselves. Only one virtual machine can affect the remaining VMs on the cloud . We need to analyse the attacks invading all these VMs so that we prevent spread of these attacks in the entire cloud. Cloud computing elucidates the concept of elastic nature to use a resource in terms of service provisioning. The cloud subscriber enjoys leasing computational resources at short notice, on either subscription or pay-per-use model and without the need for any capital expenditure into hardware. A further advantage is that the unit cost of operating a server in a large server farm being lower than in small data centers. Organizations wishing to use computational resources provided by these clouds supply virtual machine images that are running in the cloud, which allocate physical resources to virtualized operating systems and control their execution. Hence the onus is on the cloud service provider to provision the resource to support the service. To support the dynamic demand of resource provisioning without compromising the quality is a new challenge that confronts these cloud service providers. Maintaining the quality of service while lowering the cost has added a new dimension to the cloud research paradigm. Many practitioners are concerned about handling the request of these demands where controlling the over provisioning as well as under provisioning is one of the challenge. This Paper will focus mainly on three models; Security threat, resource allocation and revenue maximization models.

## 2. Related Work

In their paper, the authors highlighted that the customer could divide his data among several service providers ($SP$s) available in the market, based on his available budget[1]. Also they provided a decision for the customer, to which $SP$s he must chose to access data, with respect to data access quality of service offered by the $SP$s at the location of data retrieval[1]. This was not only to rule out the possibility of a $SP$ misusing the customers' data, breaching the privacy of data, but also could easily ensure the data availability with a better quality of service[1] . In viral marketing, a key problem is to select an initial "seed" set from the network such that the entire network adopts any behaviour given to the seed[2]. Here they introduced a method for quickly searching seed sets that scales to very large networks[2]. Their approach found a set of nodes that guarantees spreading to the entire network under the tipping model[2]. Infrastructure as a Service (IaaS) serves as the foundation layer for the other delivery models, and a lack of security in this layer will certainly affect the other delivery models, i.e., PaaS, and SaaS that are built upon IaaS layer[7]. From their point of view, the number one service or feature that was missing was security of data[7]. There were two levels of concern here. One was focused on preventing others (such as another customer) from reading private data. This was a clear and obvious concern and prominent in scenarios such as theft, or other direct malicious attack. The other was concerned with the service provider reading private data. Besides simple lack of trust of the provider themselves, it should be obvious that the service provider is not 100% immune to attacks or other malicious activity, targeted or otherwise. These two levels of concerns applied to other security issues as well, and of course were commensurate with the level of confidentiality desired[8].

They considered intruder model and requirements that need to be satisfied to provide required level of privacy. Since previous research show that crypto- graphic means cannot always provide protection (especially in long term) they proposed a trust-based privacy protection. Their approach was based on subjective logic that applied to measure/monitor level of trustworthiness of cloud service providers. They explained how users have to handle their data to minimize privacy treats in the cloud[9]. In their paper, they proposed a security metric that enables service providers and service subscribers to quantify the risks that they incurred as a result of prevailing security threats and system vulnerabilities[15]. The security metric they proposed in their paper was quantified in economic terms, thereby enabling providers and subscribers to weight these risks against rewards, and to assess the cost effectiveness of security countermeasures. Critical to the identification of threats is using a threat categorization methodology. A threat categorization such as STRIDE can be used, or the Application Security Frame (ASF) that defines threat categories such as Auditing & Logging, Authentication, Authorization, Configuration Management, Data Protection in Storage and Transit, Data Validation, Exception Management. The goal of the threat categorization is to help identify threats both from the attacker (STRIDE) and the defensive perspective (ASF). DFDs help to identify the potential threat targets from the attacker's perspective, such as data sources, processes, data flows, and interactions with users. These threats can be identified further as the roots for threat trees; there is one tree for each threat goal. From the defensive perspective, ASF categorization helps to identify the threats as weaknesses of security controls for such threats. Common threat-lists with examples can help in the identification of such threats. Use and abuse cases can illustrate how existing protective measures could be bypassed, or where a lack of such protection exists. The determination of the security risk for each threat can be determined using a value-based risk model such as DREAD or a less subjective qualitative risk model based upon general risk factors (e.g. likelihood and impact)[17]. In their paper, authors used DREAD (Damage potential, Reproducibility, Exploitability, Affected users and Discoverability) modelling.

In other terms, Cloud computing provides access to IT resources as services ranging from direct access to hardware equipment to more This work is granted by the French national project Compatible One. sophisticated applications. According to this definition, one can distinguish between three levels of Cloud services, namely infrastructure as a service (IaaS), platform as a service (PaaS), and software as a service (SaaS). By deploying large-scale data-centers, Cloud service providers (CSPs) take advantage of the economy of scale to provide virtual machines (VMs) that host a wide range of applications without any restriction on the amount of required resources. Hence, end-users can rent as many VMs as they need, while saving the cost of designing, deploying, and operating a data-center. Such an economy of scale actually benefits both parties, the CSPs and the end-users. Meanwhile, it requires the CSPs to have efficient and cost-effective datacenters. As CSPs scale their data-centers in order to keep up with the growing demand for Cloud services, the capital and operation expenditures of these data-centers increase accordingly. Hence, maximizing efficiency, cost-effectiveness and utilization of the invested infrastructure becomes foremost for CSPs. However, balancing the offered quality of service (QoS) with the CSP expectations is extremely challenging, especially for highly dynamic loads. In such a dynamic environment where end-users can join and leave the Cloud at any time, CSPs should be able to provide their clients with the required services according to a given service level agreement (SLA). Consequently, an efficient and dynamic resource allocation strategy is mandatory[11].

The proposed a theoretical framework for a secure and highly efficient multicast algorithm based upon the lattice model, to distribute the data stored in cloud to several computing nodes in a secure way, Their proposed model checked for the security violations like loss of confidentiality and integrity of data by using the lattice structure which provided them with the features of multilevel security thus made it possible to efficiently and securely transfer large amount of data stored n cloud to cluster of nodes. Few salient features of their lattice based multicast algorithm are ;Confidentiality of data during migration from cloud storage to computational nodes by enhancing the security by multilevel mechanism. Integrity of data is preserved by considering the least possibility of data modification. Within a cluster of highly computational nodes, The authors would utilize optimization techniques for efficient distribution of data to all nodes . The authors would make use of optimization ideas from multicast algorithms used in parallel distributed systems and P2P systems to achieve scalability with respect to the number of computational nodes and the amount of data .

Proposed algorithm first divides the data to download from the cloud storage service over all nodes, and then exchanges the data via a mesh overlay network . Cloud systems are based on large clusters in which nodes are densely connected, Contrary to traditional clusters, the computational and storage resources provided by clouds are fully or partly virtualized. A multicast algorithm for clouds can therefore not assume anything about the exact physical infrastructure. The network performance within clouds is dynamic. The performance of the uplink and downlink of a virtual compute node can be affected by other virtual compute nodes that are running on the same physical host. Routing changes and load balancing will also affect network performance . In today's competitive economy data is the primary asset for everyone. In cloud computing, foremost concern is about data integrity and its confidentiality[6]. The concerns are as mentioned below:

How can we ensure integrity and prevents loss of our data in cloud?

How will our data remain confidential? How would we protect privacy of our data?

How do we maintain data confidentiality and integrity where several applications running on clouds need the stored data for further processing?

Various implications are discussed regarding integrity and confidential of data.

For ensuring the integrity of the data during its storage, processing and retrieval means that it changes in responses to validated movements of data. Such concerns are relatively straightforward to address through very strong encryption mechanisms like AES, DES and many more techniques. The management can be done through common PKI infrastructure. Labels are placed on repositories encrypted with a public key that is associated with each user. The user holds the private part of the key and is authorize to decrypt the labels encrypted with the public part. This form of encrypted data in cloud is good for storage or archival but is rather costly to process. However, a new form of encryption technique, called Homomorphic Encryption [8] enables the cipher text to be processed in public cloud without decrypting it. Service providers need to ensure storage integrity against loss of non-volatile data due to failure of storage sub-system and bit rots. Distributed data coding like Erasure Coding and network coding has been studied and used extensively [9], especially for fault tolerant and highly available storage in cloud. Transport level security (TLS) measures ensure secure data transfer over networks.

Confidentiality of data has to be maintained by ensuring that it is not gained by unauthorized users. The common method of masking data of customer record confidentiality is data anonymization. So in the context of risks such as banking, health, insurance, research is being performed to better common anonymization techniques like k-anonymization with distributed anonymization [6]. The authors[8] proposed the following model:

Step-I: For any time instant t, total available volume will be identified and using an already available function the volume number is divided by 10 until we get a remainder value<10.

Step-II: The sum of the volume is identified and this creates the number of layers.

Step-III: Each of the value is identified as co-efficient of the layer value for the respective layers.

Step-IV: Once the layers are created, every layer will create an agent with inherent service-request & service-response mechanism.

Step-V: An Agent in each layer checks the co-efficient value of $x_i$ for every layer.

Step-VI: If $c_i > 9$, the agent $A_i$ will send a message to higher layer agent $A$ to add co-efficient so that $A_{i+1}$ will assign its co-efficient as $(c_{i+1} + 1)$.

If $c_i < 9$, the agent $A_i$ will send a message to lower layer agent $A_i$ - 1 to subtract co-efficient so that the Agent $A_i$ -1 will make its co-efficient as $(c_{i-1} - 1)$.

Step-VII: For $c_i > 9$, the agent $A_i$ will increase the coefficient as $(c_i + 10)$. For $c_i > 9$, the agent $A_i$ decrease the co-efficient value to $(c_i - 10)$.

The comparison takes place with value 9 always. This value is taken as a constant for static situation. 9 being the largest value in decimal is another reason behind our choice [6].

An "Agent" is an autonomous entity. The agent performs certain tasks on behalf of the user. The agent has autonomous nature, is adaptive and asynchronous. We have used an "Agent" for decision making and controlling the internal environment.

## 3. The aim of the survey paper

The aim of this survey paper is to come up with concrete knowledge on applying mathematical models on the cloud based on the proposed models:

1. Security model
2. Revenue maximization model
3. Resource allocation model

Firstly, we assume that VM 1(virtual machine 1) is the source of attack and carries viruses. Due to the sharing of resources, it affects any virtual machine on the cloud call it physical machine. Once the physical machine is affected, there is a possibility that VM 2, VM

3.........VM n be affected and process continues until the entire cloud is affected. Here, we use Las Vegas Randomized Algorithm(LVRA) which state that " you will always get a solution if there is a solution at all. The task is to figure out how to simulate the first attack successfully.

Secondly, we use Stochastic modeling to estimate the probability of outcomes within a forecast to predict what conditions might be like under different situations. The random variables are usually constrained by historical data, such as past market returns.

Finally but not limited, we need to know the situation where users are competing for resources with different financial capacities. We assume that when proposing their requests for cloud resources, all the users offer their bids at the same time and only know their own bids. The resources are allocated later based on their bid proportions.

### A. Security threat model

1. The top security threats in cloud computing

There are security threats which attack individual systems but there are also those that are common in cloud computing at large. These include among others; abuse and nefarious use of cloud computing, insecure interfaces & API's, unknown risk profile, malicious insiders, Shared technology issues, data loss or leakage and account or service hijacking.

2. Risk Mitigation

Each threat category described by STRIDE has a corresponding set of countermeasure techniques that should be used to reduce risk. STRIDE is an abbreviation which means Spoofing user identity, Tampering with data, Information disclosure , Denial of service and Elevation of privilege. These are summarized in Table below. The appropriate countermeasure depends upon the specific attack:

Table.  STRIDE Threats and Countermeasures

| Threat | Countermeasures |
|---|---|
| Spoofing user identity | <ul><li>Use strong authentication.</li><li>Do not store secrets (for example, passwords) in plaintext.</li><li>Do not pass credentials in plaintext over the wire.</li><li>Protect authentication cookies with Secure Sockets Layer (SSL).</li></ul> |

| | - Use data hashing and signing.<br>- Use digital signatures. |
|---|---|
| Tampering with data | - Use strong authorization.<br>- Use tamper-resistant protocols across communication links.<br>- Secure communication links with protocols that provide message integrity.<br>- Create secure audit trails. |
| Repudiation | - Use digital signatures.<br>- Use strong authorization |
| Information disclosure | - Use strong encryption.<br>- Secure communication links with protocols that provide message |
| Denial of service | - Validate and filter input. |
| Elevation of privilege | - Follow the principle of least privilege and use least privileged service accounts to run processes and access resources. |

## 3. Viral attack in Virtual Machine (VMs)

The way how viruses attack the human cell's body resembles that one of virtual machines on cloud. The cloud providers try to isolate virtual machines from the physical machine to maintain the security of data that may be attacked and destroy the entire system. But still the security of virtual machine is not maintained. The dynamic sharing of resources from physical server or among the virtual machines themselves will facilitate the viruses to deploy their attack from one VM to the next until the entire system is affected. The figure below shows that the viral attack may come from any point and invade the cloud which has different virtual machines.

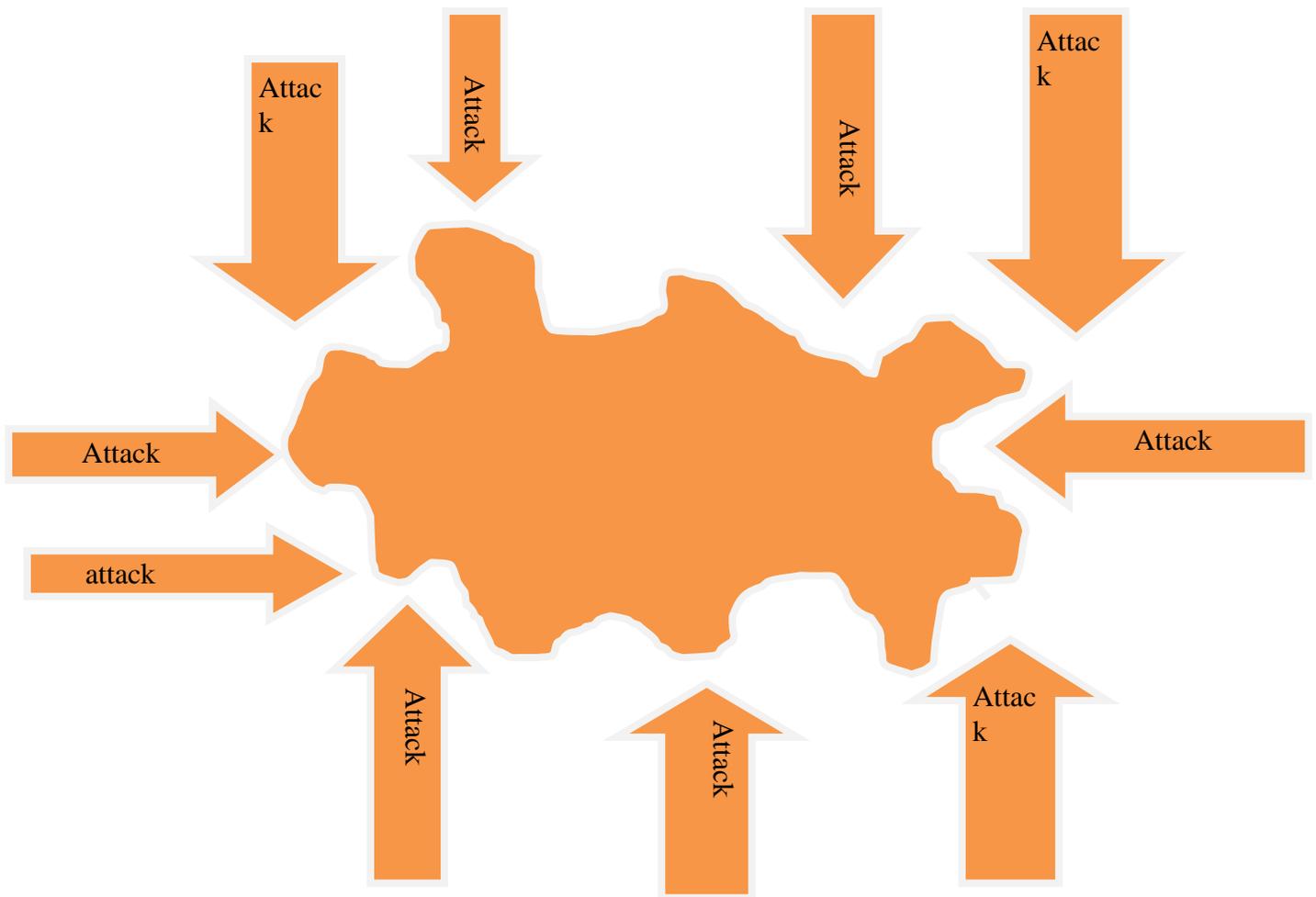

Figure 1: Viral attack VMs on the cloud

In the figure 2 below, the arrows show the different attacks to virtual machines(VMs) on the cloud. These attacks grow gradually until the entire system is affected. The cloud users can lose their data once these viral attack are spread all over the system. In this paper we propose viral model system that will prevent all these attacks. Once one virtual machine(VM) is attacked, there is a possibility that the second virtual machine(VM) may be also attacked and so on.

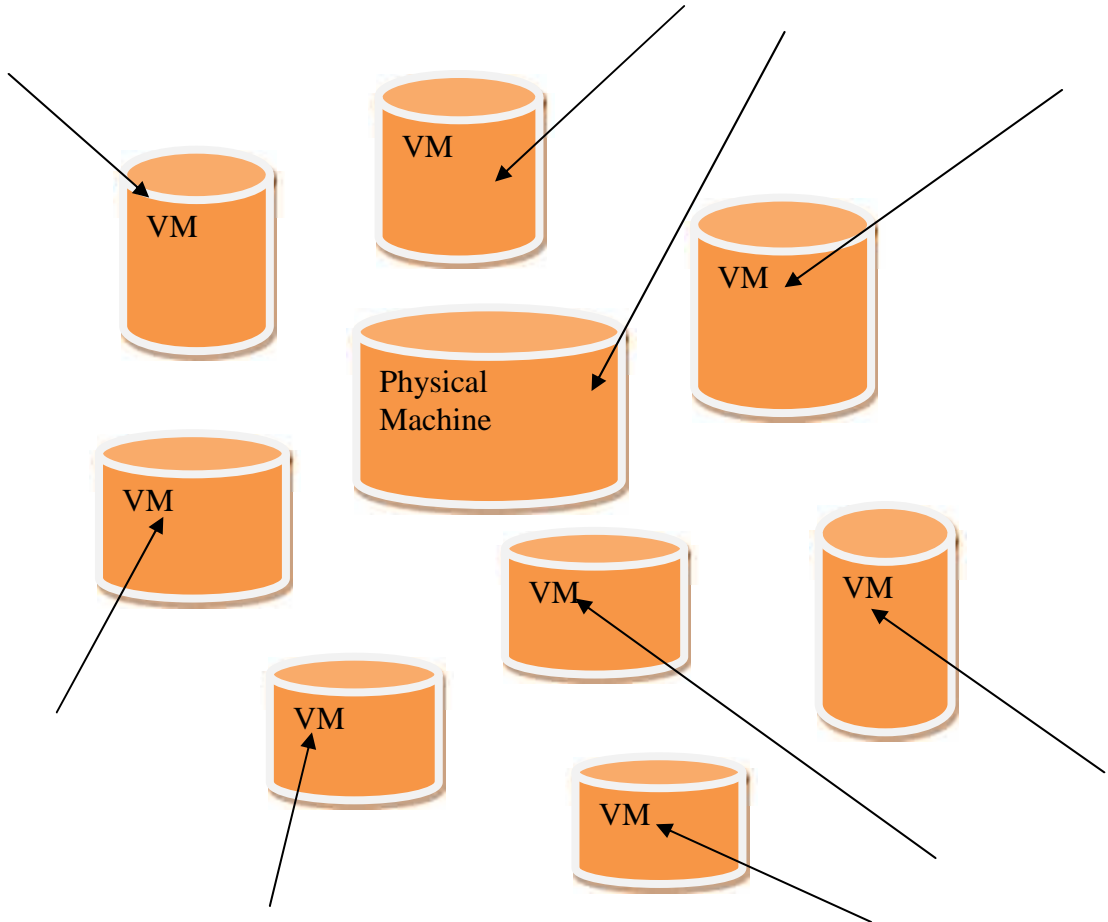

Figure 2 : Viral attack on VMs in the cloud

## 4. Spreading of Viral Attacks on Cloud

Here we assume that VM 1 is the source of attack and carries viruses. Due to the sharing of resources, it affects any machine on the cloud call it physical machine. Once the physical machine is affected, there is a possibility that VM 2, VM 3.........VM n be affected and process continues until the entire cloud is affected. This scenario is shown in the figure below taking into account of all VMs:

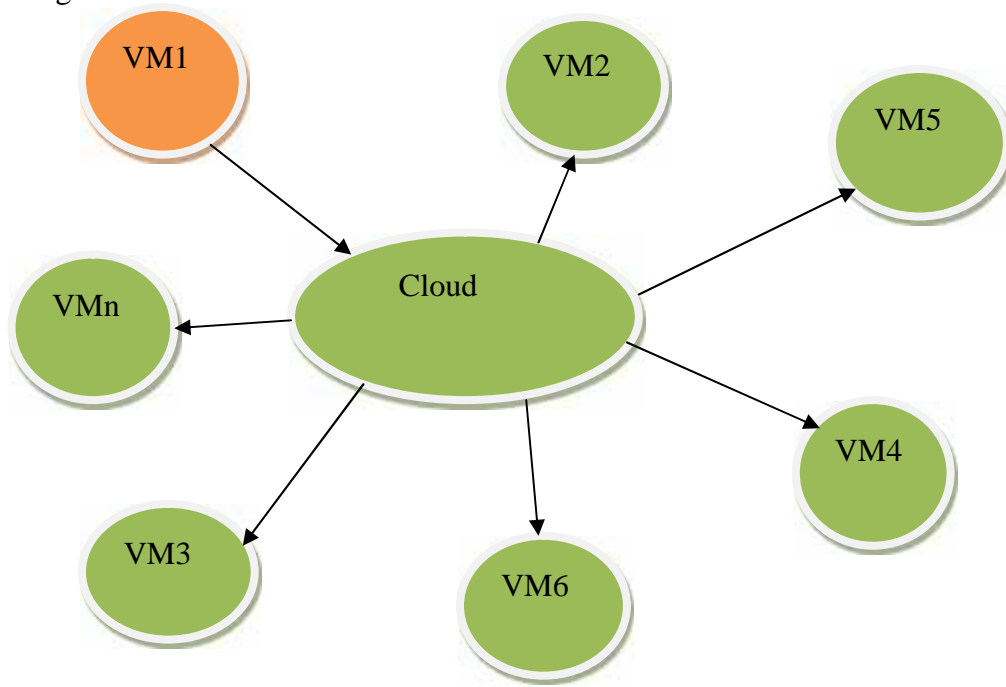

## 5. Proposed Viral model system

Assume that at least one VM is affected, virus is seeded onto other VMs. The dynamic sharing of resources among VMs allow each VM to be affected by viruses and eventually affects the entire cloud. In this model, we use Las Vegas Randomized Algorithm(LVRA) which state that " you will always get a solution if there is a solution at all. The task is to figure out how to simulate the first attack successfully. In this model, we use Random Fibonacci sequences. This can be demonstrated below:

Given that:

$$L_n = \frac{1}{2}(a_{n-1} + a_{n+1}) \qquad *$$

$$L_0 = 2, L_1 = 1$$

Initially no nodes are affected " node 1" may be affected

$a_0 = 0, a_1 = 1$ are the first pair of seeds. $L_n$ is the possible number of nodes to be affected

Now, $a_n$ is a solution (well-known) to the deterministic difference equation below:

$$a_{n+1} = a_{n-1} + a_n, n \geq 2$$

$$a_n = 0, a_1 = 1$$

from  * above, we assume that All VMs talk to each and    is the Random Fibonacci sequence.

## 6. The growth of threat attacks with respect to their percentage per year

The magnitude of attacks have increasingly observed since 1960s up to 2003 when they found out the solutions to avoid all these attacks. As you can see in the figure below, in 1990s people noticed that it was becoming was called this year " the cock's eggs". But still more have to be done as the various technologies are increasingly available and awareness of the people to use these technologies. we need to reduce on the magnitude of these attacks up to 1% or even 0 % if at all we need perfect communication otherwise it will remain a challenge to all cloud users and information society in general.

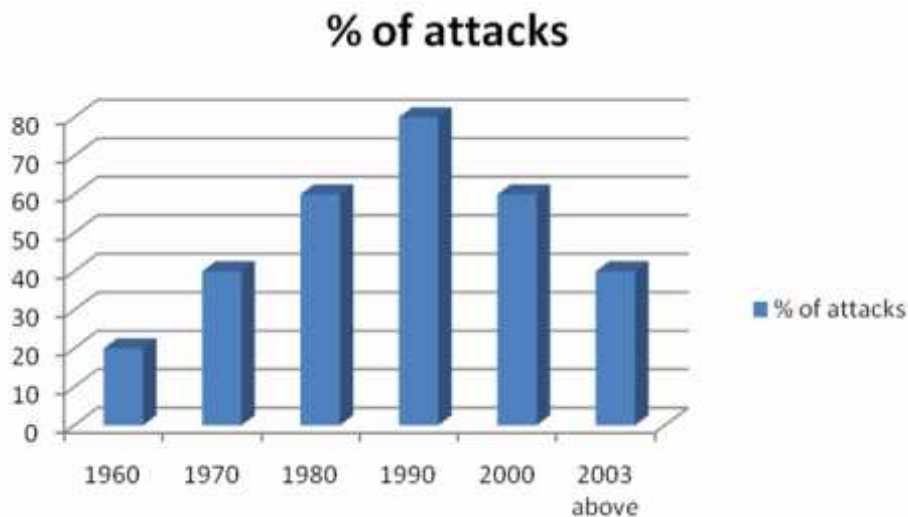

Figure 4: The percentage of threat attacks per year

**B. REVENUE MODELS**

This survey explains two important revenue maximization models which include; A stochastic and its distributions for revenue maximization. This is method of financial modeling in which one or more variables within the model are random. Stochastic modeling is for the purpose of estimating the probability of outcomes within a forecast to predict what conditions might be like under different situations. The random variables are usually constrained by historical data, such as past market returns.

**1. A stochastic revenue maximization**

 **Formulations**

The pricing problem can be formulated as follows. At the current time, the operator has $x$ [0,C] spot instances running in the system with capacity $C$. It faces a finite decision horizon t>0 to collect revenue, until it updates the demand functions f(*) and g(*).

Note here $t$ essentially indicates *how much time is left for sale*, and decreases along the time line. The provider uses a non-anticipating pricing policy $p(s)$ to maximize the expected revenue over the entire decision horizon. Let $X(s)$ denote system utilization, i.e. the number of active instances in the system at any time s [ 0,t]. A demand is realized at time $s$ if $dX(s) = 1$, and is vanished at time $s$ if $dX(s) = -1$. The pricing policy must be such that the number of active instances does not exceed the capacity $C$ at any time $s$. Author denote by $U$ the set of all such possible pricing policies that satisfy $s$ (0)

$$\int_0^s dX(m)[-x, C-x] \tag{1}$$

$$(p) \in [0,1] \forall [0,t] \tag{2}$$

Here *m* denotes time in [0, s] when *s* is given. Constraint (1) is the capacity constraint mentioned above. The existence of null prices guarantees that it can always be satisfied. Given a pricing policy u , U. The author denoted the expected revenue collected over the time period [0, t] by

$$J_u(x, t) = E_u[\int_0^t p(s)X(s)ds] \; \forall t > 0 \tag{3}$$

At the very end of the horizon when *t* = 0, the expected revenue is clearly zero for any utilization x.

$$J_u(x,0) = 0, \forall x \in [0, C]. \tag{4}$$

The provider's problem is to find a pricing policy *u* that maximizes the expected revenue generated over [0, t], denoted by $J^*(x, t)$. Equivalently, $J^*(x, t) = \sup_{u \in U} Ju(x, t)$

(5)

2. Stochastic System Model

The stochastic system model helps us analyze the conditions under which dynamic pricing outperforms FCFS. In addition, the model can also be used to calculate the optimum reservation prices and thresholds for each situation. To model this scenario the authors used a Markov chain-based approach.

The model was presented in three steps: first of all, the assumptions and an example scenario were introduced. Subsequently the different states, their definition and the calculation of their transition probabilities were described for the case of a discrete distribution.

The amount of states, however, rapidly grows as the granularity of the model increases. With a continuous distribution the state space is infinite. To avoid this, state classes which pool states with similar attributes were introduced in the third step. This helps to simplify the model representation and drastically reduce the state space. This results in a model with a reduced number of states and also allows for the introduction of continuous distributions. For models with discrete distributions the transition from the second to the third step merely shifts information from states to transition probabilities. Therefore the results of both models are the same. This leads to the following assumptions for this model.

• A1: Job arrival follows a given stochastic process.

- A2: Job ending follows a given stochastic process.
- A3: A certain amount of resources is available.
- A4: The price of a job is a random variable X with a known distribution / cumulative distribution function.
- A5: Jobs are instantly accepted or rejected.
- A6: In case of dynamic pricing if a certain utilization threshold is exceeded jobs are only accepted if their price is higher than the respective reservation price.

Resources required by each job can vary, they can be modeled as random variable with a known distribution function or using fuzzy sets if sufficient probabilistic information is not available.

To better illustrate the model authors introduced the following example scenario, which would be used for the remainder of that paper. Since authors focused on the effects different prices had on revenue, authors used a constant resource requirement for each job (20% percent of the available capacity). Job revenues follow a discrete uniform distribution from 0.5 to 1.5 with an interval of 0.2. For the classification of the states, the capacity utilization (C0-C5) and the average revenue of each job are used. The average revenue per job $r_{avg}$ would be grouped into three classes : This was shown below: $\left(R1: 0.5 <= r_{avg} <= 5/6, R2: 5/6 < r_{avg} <= 7/6, R3: 7/6 < r_{avg} <= 1.5\right)$.

## 2. Stochastic Model with discrete distribution

Intuitively states could be described by the amount of running jobs of type A and type B. As job revenues are drawn from a distribution, this characterization would lead to complex state definitions. Furthermore it would result in numerous states with different job combinations but identical utilization and revenue. To avoid that, a different definition of states is used. Each state within the model can be described by two values: the capacity used (or the number of jobs running) and the total revenue of all jobs running. So one state consists of a pair of values and the notation of capacity/revenue is sufficient to identify one state.

However each such state can be reached via different paths within the model. For example, the state "2/1.4", which represents a state with two running jobs and the total revenue of 1.4, can be obtained by three different combinations. The job combination 0.5 and 0.9, two jobs 0.7, as well as the job combination 0.9 and 0.5 lead to the same state. This state property has to be taken into account when calculating the transition probabilities. Furthermore it must be ensured that the Markov property still is fulfilled, i.e. the transition probabilities must no depend on past states. The different transitions from each state can be grouped in the same three categories. Either a job starts or ends or no start and no end occurs. A probability is assigned to each event. All probabilities sum up to one. A new jobs starts with an endogenously given probability. Each possible new job with a certain revenue has the same probability to arrive. The probability that a specific job starts is therefore given by the probability of a job start divided by the number of possible jobs. At the capacity limit no new job can be accepted and for this reason there is no state with a higher capacity and no transition to that non-existing state. In this case, the job start probability is added to the probability that no change occurs (as every additional job would be rejected).

At every state one of the already running jobs ends with the endogenously given probability. To calculate the transition probability from a given state to a state with less capacity used, it is necessary to know which kind of jobs are running. Each running job has the same probability to end. As pointed out in the example above, there are three different combinations to reach state "2/1.4". The result, if the jobs of all three combinations are summed up, is as follows: 2 times a job with revenue 0.5, 2 times a job with revenue 0.7 and 2 times a job with revenue 0.9. All in all 6 jobs. The probability, if a job ends (due to the given job end probability), that it is one of the jobs with 0.5 revenue is 2/6. Therefore, the transition probability from state "2/1.4" to state "1/0.9" (which is: a job with revenue 0.5 ends) is the product of the endogenously given job end probability and 2/6. The other transition probabilities are computed accordingly. Transition probabilities from a state to itself are given. No calculations are necessary except for the states "0/0", the capacity limit, and the overload states.

## 3. General Stochastic Model with state classes

State classes pool several states into one class. A class is defined by the capacity used and a revenue range. Each state with the capacity and the appropriate revenue is assigned to the respective class. Instead of using the total revenue of each state, the average revenue is a better measure as it allows comparing states and classes between different capacity stages.

The average revenue is given by the total revenue divided by the number of running jobs.

## 4. Optimality Conditions

Equation (5) is a stochastic dynamic programming problem. To solve it, authors could consider its Hamilton-Jacobi conditions, which are the continuous-time counterpart of the Bellman equation. Informally, consider what happens over a small interval of time $\Delta t$. Since both the arrival and departure processes are Poisson, by selecting a price $p$, the provider sees one more instance over the next $\Delta t$ with probability $f(p)\Delta t + o(\Delta t)$, one fewer instance with probability $g(p)\Delta t + o(\Delta t)$, and no change with the rest of the probability mass. By the Principle of Optimality,

$$J^*(x, t) = \sup_p [px\Delta t + o(\Delta t) + f(p)\Delta t] * J(x+1, t-\Delta t) + g(p)\Delta t * J(x-1, t-\Delta t) + (1-(f(p))$$

$$J(x+1, t-\Delta t) + g(p)\Delta t * J(x-1, t-\Delta t) + (1-(f(p)+g(p)))\Delta t)J^*(x, t-\Delta t)] \tag{6}$$

In words, with $t$ time left to the end of the horizon, the optimal expected revenue $J(x, t)$ must be equal to the realized revenue during $\Delta t$ which is simply $px\Delta t$, plus the expected value of the optimal expected revenue from the remaining time interval $t - \Delta t$, which are the remaining terms of (6). Re-arranging the terms and taking the limit as $\Delta t \to 0$, we get:

$$\partial J(x, t)/\partial t = \sup_p [px + f(p)(J^*(x+1, t) - J^*(x, t)) - g(p)(J^*(x, t) - J^*(x-1, t)) \tag{7}$$

Note that (7) holds only for $1 \leq x \leq C-1$. When $x = 0$, the provider will not see any departure over $\Delta t$, and is forced to price at 0; when $x = C$ the provider is forced to set the price to 1 to shut down the arrival process as discussed above. Thus, $p(0, t) = 0$ and $p(C, t) = 1$ in their model. Authors had the following equations:

$$J^*(0, t) = f(0)\mu t\, J(1, t-\mu t) + (1 - f(0)\mu t)J^*(0, t-\mu t) + o(\mu t),$$

$$J^*(C, t) = g(1)\Delta t * J(C-1, t-\Delta t) + (1 - g(1)\Delta t)J^*(C, t-\Delta t) + C\Delta t + o(\Delta t),$$ from which we obtain the following conditions:

$$\partial J^*(0, t)/\partial t = f(0)J^*(1, t) - J^*(0, t) \tag{8}$$

$$\partial J^*(C, t)/\partial t = C - g(1)(J^*(C, t) - J^*(C-1, t)) \tag{9}$$

## 5. Benefits of dynamic pricing for users

In this paper, authors mainly focused on the providers nevertheless, dynamic pricing is also beneficial for users of the cloud. With static pricing the provider has limited ways to control the user demand. When demand for resources (e.g. virtual machines, bandwidth, etc.) increases, the performance of the virtual machines degrades and the probability of failures increases, leading to inferior user experience. With dynamic pricing, the provider has an effective means to dynamically control the demand, and ensure the overall performance of the cloud is satisfactory for customers. Thus authors believed that dynamic pricing is also beneficial to users, from the performance point of view.

They did not explicitly model the resource contention (e.g. bandwidth, CPU, memory) and its effect on user experience of using the cloud, which is an interesting future direction of extending the work. Such an effect is in fact an important topic in our community.

Recently there has been active research on providing bandwidth guarantees and different notions of fairness to users of the cloud .They have also reached out to public cloud providers such as Microsoft Azure to comment on the practicality of dynamic pricing . They view dynamic pricing as an viable option that will be increasingly.

**C. Resource Allocation Models**

In a real-life scenario, cloud computational resources are shared among different cloud consumers who will pay for the services according to their usage of resource. Generally, the resource details are hidden from users through virtualization. Observed from user perspective, services are identical in terms of functionality and interface. However, it is not financially reasonable to provide the same QoS to the users who would like to pay more for better services. Authors' paper studied the situation where users were competing for resources with different financial capacities. They assumed that when proposing their requests for cloud resources, all the users offered their bids at the same time and only know their own bids. The resources were allocated later based on their bid proportions.

Then , authors looked at a general case as an example where cloud provider virtualized $K$ resources totally, each of which could render a specific service with a fixed finite capacity $C = [C1, C2, . . . , CK]$. These resources would be allocated to cloud users using bid proportion allocation mechanism.

Authors assumed that there are $N$ users who hold the same jobs in cloud market. Each job is composed by a set of sequent subtasks where $q_i^k$ stands for the task size. Meanwhile, every user has its own bidding function, which calculates the suitable estimated cost to purchase a resource depending on task size, priority, QoS requirement, budget and deadline. Users will bid for their subtasks according to their bidding functions. In their paper, normal distribution, which is popular in stocks market nowadays, is employed to describe the financial capability of the users:

p(Bi) * N(μ$_i$,  2) where *Bi* is the money that user would like to pay for hiring the resource for a second distributed normally with mean *μ$_i$* and variance  2. User *i*  bids for task *k* at price *bi k* that can be considered as a sample for *Bi*. The authors found that the similar scenarios were well analyzed in game theory, in which an individual's success in making choices depends on the choices of others . In their paper, they  used the equilibrium model to estimate the final state of such a competition scenario and build a market-oriented resource allocation mechanism with two types of constraints: budget and deadline. The policy-based decision model does not impose restrictions on the type of workload, demand or price distribution. The stochastic system model however required certain assumptions in order to help them calculate the expected revenue, utilization and rejection rate. Job arrival (and ending) must follow a stochastic process. Furthermore the cumulative distribution function for job prices must be known. However, these assumptions do not restrict applicability. It can be used in Infrastructure/Platform- as well as Software-as-a Service scenarios. However the precision, optimal policies and optimal configuration will vary with different scenarios. Unlike related approaches their model did not require precise demand forecasts and was not restricted to certain workload shapes. However, if such precise forecasts are available they can be incorporated (e.g. using time-in homogenous transition probabilities) to improve the results. The previous model did not account for influences the dynamic pricing might had on customers (e.g. inducing them to shift their jobs to times of lower demand). Such behaviour is inherently difficult to model and highly depends on the user base as well as the provider's position in the market. If suitable models for such consumer behaviour become available, it should be possible to integrate them into their approach. Both their decision model and their system model worked with different resource requirements for each job. However, due to space constraints and the focus on the effects of pricing they used constant or known resource requirements in this evaluation. While this is a realistic assumption for some services (e.g. Amazon EC2 with predetermined resource bundles, online storage), it cannot be assumed generally. A more detailed evaluation of the decision model with real world workloads with varying capacity can be found in this article. The policies presented in that work and used for the evaluation did not differentiate between certain types of customers (e.g. regular pay per use vs. long term service contract, key vs. regular customer). Work on client classification can be found in as well. The managerial implications are straightforward. The model supplies providers with a toolbox which can help them predict revenue, utilization of the systems as well as the probability that jobs are rejected due to insufficient capacity. The influence on revenue of changes in demand and the price distribution prevalent on the market can be quantified using their model. The proposed model can be used to identify optimal policy configuration in terms of revenue or availability for each situation and thus help provider increase their revenue significantly. The prediction of the utilization and the probability that jobs are rejected due to insufficient capacity can help providers in determining the optimal size of their infrastructure and when to extend capacity. The prediction of the probability that jobs are rejected due to insufficient capacity can further be used for determining realistic service levels that can be offered and codified in the SLAs. It can further help to predict the loss of revenue or cost incurred by improper service levels. In resources allocation, they considered two types; Static and dynamic resources allocation: This is the situation where the expected resource allocation is predictable and does not have high volatile demand .Authors were considering this as a general scenario. Cloud computing is on demand as it offers dynamic flexible resource allocation for reliable and guaranteed services in a pay-as-you-use manner to public. In cloud computing, a cloud resource consumer can request a number of cloud resources simultaneously. So there must be a provision that all resources are made available to requesting cloud resource consumers in an efficient manner to satisfy their need.

## 9. Conclusion

Cloud computing is on demand as it offers dynamic flexible resource allocation for reliable and guaranteed services in a pay-as-you-use manner to public. In cloud computing, a cloud resource consumer can request a number of cloud resources simultaneously. So there must be a provision that all resources are made available to requesting cloud resource consumers in an efficient manner to satisfy their need**.** Due to the fact that there is a pool of cloud users, it requires high security and availability through resource allocations. The cloud consumers also need to gain profit and revenue maximization.  In this paper, we discussed different security threats and countermeasures and propose threat modelling to combat these attacks. It is important to  come up with different models in which either the  providers or cloud user find it easy to use, flexibility , availability and accessibility  of the resource allocated on the cloud. It is also equally important to discuss the revenue models so that the cloud users may benefit from what they expect from cloud services including IaaS, PaaS, SaaS  and so on .The Survey considers all  these models and provides also related work on the matter. The mathematical expressions and models have been seen as one of tools that can be used to analyse and mitigate threats or attacks that may affect the entire cloud.  It is also helpful to use the mathematical models to allocate resources and determine revenue and pricing models on the cloud. The virtual machines (VMs)  can be tested  due to the sharing of resource among these VMs.